\documentclass[aoas,MSNbibl,nameyear,dvips]{arximspdf}
\usepackage{graphicx}

\doi{10.1214/11-AOAS467}
\volume{5}
\issue{3}
\pubyear{2011}
\firstpage{1893}
\lastpage{1919}

\makeatletter
\def\bptnote#1{}
\makeatother

\begin{document}
\begin{frontmatter}

\title{Efficient methods for sampling spike trains in networks of coupled neurons\thanksref{TITL}}
\runtitle{Sampling spike trains}
\thankstext{TITL}{Supported by a McKnight Scholar and NSF CAREER award, as well as Grant IIS-0904353 from the NSF.}

\begin{aug}
\author[A]{\fnms{Yuriy} \snm{Mishchenko}}
and
\author[A]{\fnms{Liam} \snm{Paninski}\corref{}\ead[label=e1]{liam@stat.columbia.edu}%
\ead[url,label=u1]{www.stat.columbia.edu/\textasciitilde liam}}

\runauthor{Y. Mishchenko and L. Paninski}

\affiliation{Columbia University}

\address[A]{Columbia University\\
1255 Amsterdam Ave, New York\\
New York 10027\\
USA\\
\printead{e1}\\
\printead{u1}} 
\end{aug}

\received{\smonth{6} \syear{2010}}
\revised{\smonth{11} \syear{2010}}

\begin{abstract}
Monte Carlo approaches have recently been proposed to quantify
connectivity in neuronal networks.  The key problem is to sample from
the conditional distribution of a single neuronal spike train, given
the activity of the other neurons in the network.  Dependencies
between neurons are usually relatively weak; however, temporal
dependencies within the spike train of a single neuron are typically
strong.  In this paper we develop several specialized
Metropolis--Hastings samplers which take advantage of this
dependency structure.  These samplers are based on two ideas: (1) an
adaptation of fast forward--backward algorithms from the theory of
hidden Markov models to take advantage of the local dependencies
inherent in spike trains, and (2)~a~first-order expansion of the
conditional likelihood which allows for efficient exact sampling in
the limit of weak coupling between neurons.  We also demonstrate that
these samplers can effectively incorporate side information, in
particular, noisy fluorescence observations in the context of
calcium-sensitive imaging experiments.  We quantify the efficiency of
these samplers in a variety of simulated experiments in which the
network parameters are closely matched to data measured in real
cortical networks, and also demonstrate the sampler applied to real
calcium imaging data.
\end{abstract}

\begin{keyword}
\kwd{Neural spike train data}
\kwd{fluorescence imaging}
\kwd{HMM}
\kwd{block-Gibbs sampling}.
\end{keyword}

\end{frontmatter}

\section{Introduction}
One of the central goals of neuroscience is to understand how the
structure of neural circuits underlies the processing of information
in the brain, and in recent years a considerable effort has been
focused on measuring neural connectivity empirically
[\citet{Shepherd03}; \citet{Bureau2004}; \citet{Briggman2006}; \citet{Hagmann2007};
\citet{SatoSvoboda07}; \citet{Smith2007}; \citet{Hagmann2008}; \citet{Luo2008};
\citet{Bohland2009}; \citet{Helmstaedter2009}].
``Functional'' approaches to this neural connectivity problem rely on
statistical analysis of neural activity observed with experimental
techniques such as multielectrode extracellular recording
[\citet{HATS98}; \citet{HARR03}; \citet{Stein04}; \citet{PAN04c};
\citet{TRUC05}; \citet{Santhanam06}; \citet{Harris07};
\citet{PILL07}]
or calcium imaging [\citet{Tsien89}; \citet{CAR03}; \citet{ImagingManual}; \citet{OHKI05}].
Although functional approaches
[\citet{BRIL88}; \citet{NYK03b}; \citet{NYK05}; \citet{OKA05}; \citet{LAKSH06};
\citet{Rigat06}; \citet{NYK06}; \citet{YuSahani06}; \citet{KP06}]
have not yet been demonstrated to directly yield the true physical
structure of a neural circuit, the estimates obtained via this type of
analysis play an increasingly important role in attempts to understand
information processing in neural circuits
[\citet{Bureau2004}; \citet{SatoSvoboda07}; \citet{PILL07}; \citet{Mishchenko2010}].

Perhaps the biggest challenge for inferring neural connectivity from
functional data---and indeed in network analysis more generally---is
the presence of hidden nodes which are not observed directly
[\citeauthor{Nykamp05} (\citeyear{Nykamp05}, \citeyear{NYK06});
\citet{KP06}; \citet{Vidne08}; \citet{PL07};
\citet{Vakorin09}].  Despite swift progress in simultaneously
recording activity in massive populations of neurons, it is still
beyond the reach of current technology to monitor a complete set of
neurons that provide the presynaptic inputs even for a single neuron
[though see, e.g.,\ \citet{Petreanu09} for some recent progress in
this direction].  Since estimation of functional connectivity relies on
the analysis of the inputs to target neurons in relation to their
observed spiking activity, the inability to monitor all inputs can
result in persistent errors in the connectivity estimation due to model
misspecification.  Developing a principled and robust approach for
incorporating such unobserved neurons, whose spike trains are unknown
and constitute hidden or latent variables, is an area of active
research in connectivity analysis [\citet{NYK06};
\citet{PL07}; \citet{Vidne08}; \citet{Vakorin09}].

Incorporating these latent variables into connectivity estimation is a
challenging task.  The standard approach for estimating the
connectivity requires us to compute the likelihood of the observed
neural activity given an estimate of the underlying connectivity.
Computing this likelihood, however, requires us to integrate out the
probability distribution over the activity of all hidden neurons.
This latent activity variable will typically have very high
dimensionality, making any direct integration methods infeasible.

Thus, it is natural to turn to Markov chain Monte Carlo (MCMC)
approaches here [\citet{Rigat06}].  The best design of such an MCMC
sampler is not at all obvious in this setting, since it may be
necessary to develop sophisticated proposal densities to capture the
dependence of the hidden spike trains on the observed spiking data in
order to guarantee a reasonable proposal acceptance rate [\citet{PL07}].
For example, the simplest Gibbs sampling approaches may not perform
well given the strong dependencies between adjacent spiking timebins
that are inherent to neural activity.

In this paper we develop Metropolis--Hastings (MH) algorithms for
efficiently sampling from the probability distribution over the
activity of hidden neurons. We derive a proposal density that is
asymptotically correct in the limit of weak interneuronal couplings,
and demonstrate the utility of this proposal density on simulated
networks of neurons with neurophysiologically feasible parameters
[\citet{Sayer1990}; \citet{Abeles91}; \citet{Braitenberg1998};
\citet{Urquijo2000}; \citet{Song2005}; \citet{Lefort2009}].
We also consider a hybrid MH strategy based on fast hidden Markov
model (HMM) sampling approaches that can be exploited to extend the
range of applicability of the basic algorithm to more strongly coupled
networks of neurons.  In each case, the resulting MCMC chain mixes
quickly and each step of the chain may be computed quickly.

Of special interest is the problem of sampling from the probability
distribution over unknown spike trains when a neuron is indirectly
observed using calcium imaging (instead of electrical recordings)
[\citet{Tsien89}; \citet{ImagingManual}; \citet{CAR03};
\citet{OHKI05}; \citet{Vogelstein2009}].  This
problem can be naturally related to the sampling problem described
above. We develop an approach to incorporate calcium fluorescence
imaging observations directly into our proposal density.  This allows
us to efficiently sample from the probability distribution over the
activity of multiple hidden neurons given calcium fluorescence
observations, improving on the methods introduced in
\citet{Mishchenko2010}.\looseness=-1

\section{Methods}

\subsection{Model definition}\label{secIIA}

We model the activity of individual neurons with a~discrete-time
generalized linear model (GLM)
[\citet{BRIL88}; \citet{CSK88}; \citet{BRIL92}; \citet{PG00};
\citet{PAN03d}; \citet{PAN04c}; \citet{Rigat06}; \citet{TRUC05};
\citet{NYK06}; \citet{KP06}; \citet{PILL07}; \citet{Vidne08}; \citet{Stevenson2009}]:
\begin{eqnarray}\label{eqn:genmodel}
n_i(t) &\sim& \operatorname{Bernoulli} [f(J_i(t))\Delta],
\nonumber\\[-8pt]\\[-8pt]
J_i(t) &=& b_i(t) + \sum _{j=1}^N\sum _{t'<t} w_{ij}(t-t')n_j(t'),\nonumber
\end{eqnarray}
where the spike indicator function for neuron $i$, $n_i(t)\in
\{0,1\}$, is defined on time bins $t$ of size $\Delta$. Connectivity
between neurons is described via the ``connectivity matrix,''
$w_{ij}(t)$; the self-terms $w_{ii}(t-t')$ describe refractory effects
(there is a minimal ``refractory'' interspike interval of a
millisecond or two in most neurons, e.g.), and $w_{ij}(t-t')$
represents the statistical effect of a spike in neuron $j$ at time
$t'$ upon the spiking rate of neuron $i$ at time $t$.  $N$ is the
number of neurons in the neural population, including hidden and
observed neurons, and $b_i(t)$ denotes a baseline driving term (which
might depend on some observed covariates such as a sensory stimulus).
We assume that the maximal firing frequency $\Delta^{-1}$ is much
larger than the typical spiking rate, $\Delta^{-1} \gg E[n_i(t)]$, in
which case the Bernoulli spiking model in (\ref{eqn:genmodel}) is
closely related to models in which $n_i(t)$ is drawn from a Poisson
distribution at every time step (see the references above for a number
of examples).


In the presence of fluorescence observations from a calcium-sensitive
indicator dye (the calcium imaging setting), the model
(\ref{eqn:genmodel}) is supplemented by two variables per neuron $i$:
the intracellular calcium concentration $C_i(t)$ (which is not
observed directly) and the fluorescence observations $F_i(t)$.  We
model these terms according to a hidden Markov model governed by a~simple
driven autoregressive process
[\citet{Vogelstein2009}; \citet{Mishchenko2010}],
\begin{eqnarray}\label{eqn:genmodelF}
C_i(t) &=& C_i(t-\Delta) - \frac{\Delta} {\tau^c_i}
\bigl(C_i(t-\Delta)-C_i^b\bigr) + A_i n_i(t), \nonumber\\[-8pt]\\[-8pt]
F_i(t) &\sim& \mathcal{N}[S(C_i(t)),V(C_i(t))].\nonumber
\end{eqnarray}
Thus, under nonspiking conditions, $C_i(t)$ is set to the baseline
level of $C_i^b$. Whenever the neuron fires a spike, $n_i(t)=1$, the
calcium variable $C_i(t)$ jumps by a fixed amount $A_i$, and
subsequently decays with time constant~$\tau^c_i$; $\tau^c_i$ is on
the order of hundreds of milliseconds in the cases of most interest
here.  The fluorescence signal $F_i(t)$ corresponds to the count of
photons collected at the detector per neuron per imaging frame. This
photon count may be modeled with approximately Gaussian statistics
(Poisson photon count models are also tractable in this context
[\citet{Vogelstein2009}], though we will not pursue this detail here),
with the mean given by a saturating Hill-type function
$S(C)=C/(C+K_d)$ \citet{Yasuda2004} and the variance~$V(C)$ scaling
with the mean.  See \citet{Vogelstein2009} for full details and further
discussion.  Note that observations of calcium-indicator fluorescence
are typically performed at a low ``frame-rate,'' $\mathit{FR}$, measured in
frames per second.  Thus, we will restrict our attention to the case
$\mathit{FR} < \Delta^{-1}$, that is, $F_i(t)$ observations are available only
every few timesteps $\Delta$.  In addition, it will be useful to
define an effective SNR, following \citet{Mishchenko2010}, as
\begin{equation}\label{eqn:eSNR}\label{eq:eSNR}
e\mathit{SNR} = \frac{E[F_i(t)-F_i(t-\Delta)  |  n_i(t)=1]}
{E[(F_i(t)-F_i(t-\Delta))^2/2  |  n_i(t)=0]^{1/2}},
\end{equation}
that is, the size of a spike-driven fluorescence jump divided by a rough
measure of the standard deviation of the baseline fluorescence.

In \citet{Mishchenko2010} we introduced an expecta\-tion-maximization
approach for estimating the model parameters given calcium
fluorescence data; \citet{PL07} discuss a related approach for fitting
the model given spiking data recorded from extracellular electrodes.
The maximization step in this context is often quite tractable,
involving a separable convex optimization problem
[\citet{SB03}; \citet{PAN04c}; \citet{KP06}; \citet{Escola07}].  The expectation step is much more
difficult; analytical approaches are often infeasible.  Monte Carlo
solutions to this problem require us to obtain samples from the
probability distribution over the activity of all hidden neurons.
This problem is the focus of the present paper.  In the context of the
expectation maximization framework, we assume therefore that an
estimate of the model parameters $\theta$ is available, and the
problem is to obtain a~sample from $P(\mathbf{n}_{\mathit{hidden}}|\mathbf{n}_{\mathit{observed}};\bolds{\theta})$
(in the case that the spike trains of a~subset of neurons is observed directly),
or $P(\mathbf{n}_{\mathit{hidden}}|\mathbf{F}_{\mathit{observed}};\bolds{\theta})$
(in the case that spike trains are
observed indirectly via calcium fluorescence measurements).  Here and
throughout we use bold notation for vectors, that is, $\mathbf{n}(t)=\{n_i(t),i=1,\ldots,N\}$
and $\mathbf{n}=\{n_i(t),i=1,\ldots,N;t\in
[0,T]\}$; $\theta$ denotes the set of all parameters in the model,
including $\mathbf{b}, \mathbf{w}, \bolds{\tau}^c$, and $\mathbf{A}$.  In cases where
no confusion is possible below we will suppress the dependence on
$\theta$.

\subsection{A block-wise Gibbs approach for sampling from the
  distribution over activity of hidden neurons} \label{secIIAB}

In \citet{Mishchenko2010}, we noted that in order to sample from the
desired joint distribution over the activity of all hidden neurons,
$P(\mathbf{n}_{\mathit{hidden}}|\cdot)$, it is sufficient to be able to efficiently
sample sequentially from the conditional distribution over one hidden
neuron given all of the other hidden neurons, $P(\mathbf{n}_i | \mathbf{n}_{\setminus i} ; \cdot)$: if this is possible, then a sample from the
full joint distribution, $P(\mathbf{n}_{\mathit{hidden}}|\cdot)$, can be obtained
using a block-wise Gibbs algorithm.  This blockwise Gibbs approach
makes sense in this context because the connectivity weights $w_{ij},
i \neq j$, are relatively weak in many neural contexts; for example,
synaptic strengths between cortical neurons are typically fairly small
(see the discussion in Section \ref{secIIF} below).  However,
dependencies between the spike indicator variables $\{n_i(s),n_i(t)\}$
within a single neuron may be quite large if the time gap $|s-t|$ is
small, since the self-terms $w_{ii}$ are typically strong over small
timescales; for example, as mentioned above, after every spike a
neuron will enter a refractory state during which it is unable to
spike again for some time period.  See \citet{Taro08} for further
discussion.

Thus, we will focus below exclusively on the single-neuron conditional
sampling problem, $\mathbf{n}_i\sim P(\mathbf{n}_i|\mathbf{n}_{\setminus
i};\cdot)$.  In the next couple sections we will discuss methods for
sampling from $P(\mathbf{n}_i | \mathbf{n}_{\setminus i})$; in Section
\ref{sec:sampl-ca} we will discuss adaptations of these methods for
incorporating calcium fluorescence observations in $P(\mathbf{n}_i | \mathbf{n}_{\setminus i}, \mathbf{F}_i)$.

\begin{figure}

\includegraphics{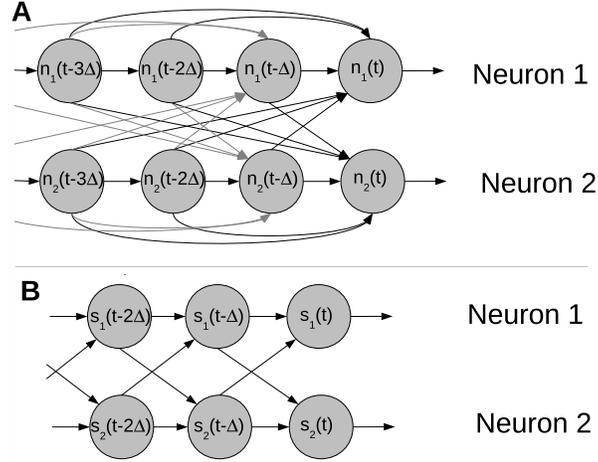}

\caption{Graphical model describing \textup{(A)} the $K$-order Markov model for
neural dynamics in terms of the instantaneous spikes, $n_i(t)$, and
\textup{(B)} a simpler Markov model in terms of the grouped states
$s_i(t)=\{n_i(t'),t-K\Delta < t' \leq t \}$.}
\label{fig:gmodel}
\end{figure}

\subsection{Exact sampling via the hidden Markov model forward--backward
procedure} \label{secIIB}

In some special cases we can solve the single-neuron conditional
sampling problem quite explicitly.  Specifically, if the support of
each of the coupling terms $w_{ij}(t)$ is contained in the interval
$[0,K \Delta]$ for some sufficiently small number of time steps $K$,
then we can employ standard hidden Markov model methods to sample from
the desired conditional distribution exactly.  The key is to note that
in this case equation (\ref{eqn:genmodel}) defines a $K$th-order Markov
process; thus, it is convenient to apply the standard substitution
$s_i(t)=\{n_i(t'),t-K\Delta < t' \leq t \}$.  Here $s_i(t)$ denotes
the ``state'' of neuron $i$, which is completely characterized by the
binary $K$-string [e.g., $s_i(t)=\mbox{``}010011100\ldots\mbox{''}$] describing its
spiking over the $K$ most recent time bins.  Clearly, the size of the
state-space is $2^K$.  See Figure~\ref{fig:gmodel} for an illustration.

To exploit this Markov structure, note that we may write the
conditional distributions $P[s_i(t) | s_i(t-\Delta), \mathbf{s}_{\setminus i}(t-\Delta)]$ and $P[\mathbf{s}_{\setminus i}(t+\Delta) |
\mathbf{s}_{\setminus i}(t), s_i(t)]$ explicitly using model
(\ref{eqn:genmodel}), and, in fact, it is easy to see that the system
\begin{eqnarray}\label{HMM1.1}
s_i(t) &\sim& P[s_i(t) | s_i(t-\Delta), \mathbf{s}_{\setminus
i}(t-\Delta)], \nonumber\\[-8pt]\\[-8pt]
\mathbf{s}_{\setminus i}(t+\Delta) &\sim& P[\mathbf{s}_{\setminus i}(t+\Delta) | \mathbf{s}_{\setminus i}(t), s_i(t)]\nonumber
\end{eqnarray}
forms an autoregressive hidden Markov model [\citet{Rabiner89}], with
$\mathbf{s}_{\setminus i}(t+\Delta)$ playing the role of the observation
at time $t$ (Figure~\ref{fig:gmodel}).  Thus, if $K$ is not too large,
standard finite hidden Markov model (HMM) techniques can be applied to
obtain the desired spike train samples.  Specifically, we can use the
standard filter forward-sample backward algorithm
[\citet{Rabiner89}; \citet{KunschIonChannels01}] to obtain samples from $P(\mathbf{s}_i | \mathbf{s}_{\setminus i})$, from which we can easily transform
back to obtain the desired sample $\mathbf{n}_i$.  This requires $O(T2^K)$
computation time and $O(2^K)$ memory, since the transition matrix
$P[s_i(t) | s_i(t-\Delta), \mathbf{s}_{\setminus i}(t-\Delta)]$ is sparse
[due to the redundant definition of the state variable $s_i(t)$], so
the matrix--vector multiplication required in the transition step
scales linearly with the number of states, instead of quadratically,
as would be seen in the general case.  Note, for clarity, that in the
above discussion we are assuming that the coupling terms $w_{ij}(t)$
are known (or have been estimated in a separate step), and therefore
$K$ is also known, directly from the maximal support of the couplings
$w_{ij}(t)$; thus, we do not apply any model selection or estimation
techniques for choosing $K$ here.

\subsection{Efficient sampling in the weak-coupling limit} \label{secIIC}

Clearly, the HMM strategy described above will become ineffective if
$K$ becomes larger than about $10$ or so, due to the exponential
growth of the Markov state-space.  Therefore, it is natural to seek
more general alternate methods.  Recall that the interneuronal
coupling terms $w_{ij}, i \neq j$, are small.  Thus, it seems promising
to examine the conditional spike train  probability in the limit of
weak coupling ($w_{ij} \to 0$), in the hope that a good sampling
method might suggest itself.

The log-likelihood of the hidden spike train $\mathbf{n}_i$ can be written as
\begin{eqnarray}\label{eqn:origprob}
\qquad \log P(\mathbf{n}_i|\mathbf{n}_{\setminus i})&=&\log P(\mathbf{n}_i,\mathbf{n}_{\setminus i}) + \mbox{const.} \nonumber\\[-8pt]\\[-8pt]
&=& \sum_i\sum_t n_i(t) \log
f[J_i(t)] - f[J_i(t)] \Delta - n_i(t)! + \mbox{const.};\nonumber
\end{eqnarray}
for notational simplicity, we have used a Poisson model for $n_i(t)$
given the past spike trains, but it turns out that we can use any
exponential family with linear sufficient statistics in the following
computations; see the \hyperref[sec:append-sampl-from]{Appendix} for details.  Now, if we make
the abbreviation
\begin{equation}\label{eqn:lagcur}
J^-_i(t) = \sum_{s<t} \sum_j w_{ij}(t-s) n_j(s),
\end{equation}
we can easily expand to first order in $w_{ij}$:
\begin{eqnarray}\label{eqn:corfull}
\hspace*{-4pt}&&  \log P(\mathbf{n}_i, \mathbf{n}_{\setminus i})\nonumber\\
\hspace*{-4pt}&&\qquad = \sum_t
n_i(t) \log f[J_i(t)] - f[J_i(t)] \Delta - n_i(t)! \nonumber \\
\hspace*{-4pt}&&\qquad \quad {}+
\sum_{j \neq i} \sum_t n_j(t) \log f [b_j(t) + J^-_j(t) ] -
f [b_j(t) + J^-_j(t) ] \Delta + \mbox{const.} \nonumber \\
\hspace*{-4pt}&&\qquad =
\sum_t n_i(t) \log f[J_i(t)] - f[J_i(t)] \Delta - n_i(t)! \nonumber \\
\hspace*{-4pt}&&\qquad \quad {}+ \sum_{j \neq i} \sum_t n_j(t) \log f [b_j(t) ] - f [b_j(t)] \Delta\nonumber \\
\hspace*{-4pt}&&\qquad \quad \hphantom{{}+ \sum_{j \neq i} \sum_t }
{}+  \biggl( n_j(t) \frac {f'[b_j(t)]} {f[b_j(t)]} - f'[b_j(t)] \Delta
 \biggr) J^-_j(t) + o(w) + \mbox{const.} \nonumber \\
\hspace*{-4pt} &&\qquad = \sum_t n_i(t) \log
f[J_i(t)] - f[J_i(t)] \Delta - n_i(t)!  \nonumber\\
\hspace*{-4pt}&&\qquad \quad {}+ \sum_{j \neq
i} \sum_t  \biggl( n_j(t) \frac {f'[b_j(t)]} {f[b_j(t)]} - f'[b_j(t)]
\Delta  \biggr) \sum_{s<t} \sum_k w_{jk}(t-s) n_k(s)\nonumber \\
\hspace*{-4pt}&&\qquad \quad {} + o(w) +
\mbox{const.} \nonumber \\
\hspace*{-4pt}&&\qquad = \sum_t n_i(t) \log f[J_i(t)] - f[J_i(t)]
\Delta - n_i(t)!  \\
\hspace*{-4pt}&&\qquad \quad {}+ \sum_{j \neq i} \sum_t \sum_{s>t}
 \biggl( n_j(s) \frac {f'[b_j(s)]} {f[b_j(s)]} - f'[b_j(s)] \Delta
 \biggr) w_{ji}(s-t) n_i(t) \nonumber \\
\hspace*{-4pt}&&\qquad \quad {} + o(w) + \mbox{const.} \nonumber \\
\hspace*{-4pt}&&\qquad = \sum_t
n_i(t) \biggl\{ \log f [J_i(t)] + \sum_{j \neq i} \sum_{s>t}
\frac {f'[b_j(s)]} {f[b_j(s)]} w_{ji}(s-t)  [n_j(s) -
f[b_j(s)] \Delta  ]  \biggr\} \nonumber \\
\hspace*{-4pt}&&\qquad \quad\hphantom{\sum_t} {}- n_i(t)! + o(w) +
\mbox{const.},\nonumber
\end{eqnarray}
where we have used the definition of $J^-_j(t)$ in the third equality,
rearranged sums (and changed variables) over $t$ and $s$ in the fourth
equality, regathered terms in the final equality, and retained only
terms involving a factor of~$n_i(t)$ throughout.

Thus, if we note the resemblance between equation~(\ref{eqn:corfull}) and
the Poisson log-probability, we see that one attractive approach is to
sample from a spike train proposal $n_i(t)$ with log-rate equal to the
term within brackets,
\begin{equation}
\log f [J_i(t)] + \sum _{j \neq i} \sum_{s>t} \frac {f'[b_j(s)]}
{f[b_j(s)]} w_{ji}(s-t)  \bigl[n_j(s) - f[b_j(s)] \Delta  \bigr];
\end{equation}
since we are conditioning on $\{n_j(t)\}$ (i.e., these terms are
assumed fixed), and there are no $n_i(s)$ terms affecting the
proposed rate of $n_i(t)$ for $s>t$, it is straightforward to sample
$n_i(t)$ recursively forward according to this rate, and then use
Metropolis--Hastings to compute the required acceptance probability and
obtain a sample from the desired distribution.

We note that this approach is conceptually quite similar to that of
\citet{PL07}, who proposed sampling recursively from a process of the
form
\begin{eqnarray}\label{eqn:PLproposal}
n_i(t) &\sim& \mbox{Poisson}[f[J^{\mathit{PL}}_i(t)]\Delta],
\nonumber\\[-8pt]\\[-8pt]
J^{\mathit{PL}}_i(t) &=& b^{\mathit{PL}}_i + \sum_{j}\sum_{s<t} w^{\mathit{PL}}_{ij}(t-s)n_j(s) +
\sum_{j \neq i} \sum _{s>t} w^{\mathit{PL}}_{ij}(t-s)n_i(s).\nonumber
\end{eqnarray}
\citet{PL07} discussed a somewhat computationally-intensive procedure
in which the parameters $b_i^{\mathit{PL}}$ and $w_{ij}^{\mathit{PL}}$ are reoptimized
iteratively via a maximum pseudolikelihood procedure.  If we reverse
the logic we used to obtain equation (\ref{eqn:corfull}), it is clear
that our approach simply uses an ``effective input''
\begin{eqnarray}\label{eqn:weakcur}
\tilde J_i(t) &=&
b_i(t) + J^-_i(t)\nonumber\\[-8pt]\\[-8pt]
&& {} + \sum_{j\neq i} \sum_{s>t} \frac{f[b_i(t)]} {f'[b_i(t)]}
\frac{f'[b_j(s)]} {f[b_j(s)]} w_{ji}(s-t)  \bigl[n_j(s)- f[b_j(s)]
  \Delta  \bigr]\nonumber
\end{eqnarray}
in place of $J_i^{\mathit{PL}}(t)$ above.  Further, in the special case of an
exponential nonlinearity, $f(J) = \exp(J)$, all pre-factors in
equation (\ref{eqn:weakcur}) cancel out, leading to an expression that
resembles equation~(\ref{eqn:PLproposal}) quite closely:
\begin{eqnarray}\label{eqn:weakcursimple}
\tilde J_i(t) &=& b_i(t) + \sum_{j} \sum _{s<t} w_{ij}(t-s)n_j(s)\nonumber\\[-8pt]\\[-8pt]
&&{} + \sum
_{j\neq i} \sum _{s>t} w_{ji}(s-t)  \bigl[n_j(s)- f[b_j(s)]\Delta \bigr].\nonumber
\end{eqnarray}
Thus, in the weak-coupling limit $w \to 0$, and ignoring
self-interaction\break terms~$w_{ii}$, we see that the optimal
``back-inputs'' $w^{\mathit{PL}}_{ij}(t-s), s>t$, from \citet{PL07} may be
analytically identified as the time- and index-reversed original
forward couplings, $w_{ji}(-t)$.

\vspace*{3pt}
\subsection{Hybrid HMM-Metropolis--Hastings approaches}\label{secIID}

So far we have developed two methods for sampling from $P(\mathbf{n}_i |
\mathbf{n}_{\setminus i})$: the HMM method (Section \ref{secIIB}) is
exact but becomes inefficient when $K$ is large, while the
weak-coupling method (Section \ref{secIIC}) becomes inefficient when
the coupling terms~$w$ become large.  What is needed is a hybrid
approach that combines the strengths of these two methods.  The key is
that the cross-coupling terms~$w_{ij}$ are typically fairly weak (as
emphasized above), and the self-coupling\break terms~$w_{ii}(t)$ are large
only for a small number of time delays $t$.

To take advantage of this special structure, we begin by constructing
a~truncated HMM that retains all coupling terms $w_{ij}(t)$ for $t$ up
to some maximal delay $t_{\max}$, where $t_{\max}$ is chosen to keep the
size of the state-space acceptably small but large enough so that the
coupling terms are captured to an acceptable degree.  Then we include
the discarded coupling terms (i.e., terms at lags longer than
$t_{\max}$) via the first-order approximation.  More concretely, denote
the conditional probability of $\{n_i(t)\}$ under the truncated HMM as
\begin{equation}\label{eqn:truncHMM}
\prod_t P_{hmm}  \bigl(s_i(t) | s_i(t-\Delta) ;\mathbf{n}_{\setminus i}
 \bigr)
\end{equation}
[recall the correspondence between the $K$-string state variable
$s_i(t)$ and the binary spiking variable $n_i(t)$].  Note that this
forms an inhomogeneous Markov chain in the grouped state variables
$s_i(t)$, and an inhomogeneous ($t_{\max}/\Delta$)-Markov chain in
$n_i(t)$, as discussed in Section \ref{secIIB}.  Now we want to
incorporate the discarded coupling terms up to the first order: we simply
form the product
\begin{eqnarray}
\label{HMMMH6}
\qquad &&\prod_t P_{hmm}  \bigl(s_i(t) | s_i(t-\Delta); \mathbf{s}_{\setminus i}
 \bigr) \nonumber\\[-8pt]\\[-8pt]
 \qquad &&\qquad {}\times\exp  \biggl( n_{i}(t)\sum _{j\neq i} \sum_{t'>t_{\max}}
w_{ji}(t'-t)\frac{f'(b_j(t'))}{f(b_j(t'))}  [n_j(t') -
f(b_j(t'))\Delta ]  \biggr).\nonumber
\end{eqnarray}
We use this simple product form here in an analogy to the HMM case, in
which we condition on observations by simply forming the (normalized)
product of the prior distribution on state variables and the
likelihood of the observations given the state variables; in this
case, the first term plays the role of the prior over the state
variables $s_i(t)$ and the second term corresponds to the
pseudo-observations represented by the weak coupling terms
incorporating the observed firing of the other neurons $n_j$.
Although, to be clear, this joint probability does not correspond
rigorously to any small-parameter expansion of $P(\mathbf{n}_i|\mathbf{n}_{\setminus i})$, we might expect that this hybrid may outperform
alternative strategies such as using only the truncated HMM, given by
equation~(\ref{eqn:truncHMM}), or only the weak-coupling correction, given
by equation~(\ref{HMMMH6}), as we will see in the \hyperref[res]{Results} section below
(see especially Figure~\ref{fig:combinedhyHMMMH}).  Since this product
form for the joint probability is structurally equivalent to the
conditional probability of an HMM in the state variables $s_i(t)$---more precisely, the graphical model [\citet{JORD99}] corresponding to the
above expression is a chain in terms of $s_i(t)$---we can employ the
forward--backward procedure to sample from this proposal, and then
compute the Metropolis--Hastings acceptance probability to obtain
samples from the desired conditional $P(\mathbf{n}_i | \mathbf{n}_{\setminus
i})$.

Note that the MH acceptance probability will decrease as a function of
the dimensionality $T$ of the desired spike train and of the size of
the discarded weights $w_{ij}(t), t>t_{\max}$, since for large $w$ the
proposal density discussed above will approximate the target density
less accurately.  In general, it is necessary to employ a blockwise
approach: that is, we update the spike train $\{n_i(t)\}$ in blocks whose
length is chosen to be small enough that the MH acceptance probability
is sufficiently high, but large enough so that the total correlations
between blocks are small and the overall chain mixes quickly.  We will
examine these trade-offs in more quantitative detail in the \hyperref[res]{Results}
section below.

\subsection{Efficient sampling given calcium fluorescence imaging
  observations} \label{sec:sampl-ca}

In this section we turn our attention to the problem of sampling from
the fluorescent-conditional distribution $\mathbf{n}_i \sim P(\mathbf{n}_i|\mathbf{n}_{\setminus i},\mathbf{F}_i)$.  In principle, we could apply
the same basic approach as before, exploiting the HMM structure of
equations (\ref{eqn:genmodel})--(\ref{eqn:genmodelF}) in the variables
$\{s_i(t),C_i(t)\}$; however, the state-space for the calcium variable
$C_i(t)$ is continuous (instead of discrete), requiring us to adapt
our methods somewhat.

We will briefly mention two alternative methods before introducing the
novel approach that is the focus of this section.  First, in
\citet{Mishchenko2010} we introduced a sampler based on a technique
from \citet{NBR03}.  This method requires drawing a rather large
auxiliary sample from the continuous $C_i(t)$ state space and then
employing a modified forward--backward approach to obtain spike train
samples.  The techniques we will discuss below do not require such an
auxiliary sample, and are therefore significantly more computationally
efficient; we will not compare these methods further here.  Second,
standard pointwise Gibbs sampling is fairly straightforward in this
setting: we write $P(\mathbf{F}_i,\mathbf{n})=P(\mathbf{F}_i|\mathbf{n}_i)P(\mathbf{n})$, and then note that both $P(n_i(t)|\mathbf{n}_{\setminus i,t})$ and
$P(\mathbf{F}_i | \mathbf{n}_i)$ can be computed easily as a function of
$n_i(t)$.  [To compute the latter quantity, note from
equation~(\ref{eqn:genmodelF}) that $n_i(t)$ only affects the values
of $F_i(t')$ appreciably for $t<t'< n_c \tau^c_i$, for a~suitably
large number of time constants $n_c$.]  We will discuss the
performance of the Gibbs approach further below.

Now we turn to the main theme of this section: to adapt the MH-based
blockwise approach discussed above, it is essential to develop a
proposal density that efficiently incorporates the observed
fluorescence data, in order to ensure reasonable acceptance rates.  We
use a filter-backward-and-sample-forward approach.  The basic idea is
to compute, via a backward recursion, the conditional future
observation density $P(F_i(t\dvtx T)|s_i(t),C_i(t);\mathbf{n}_{\setminus i})$
given the current state $(s_i(t),C_i(t))$, for all $t \leq T$.
Then, we may easily sample from a proposal of the form
\begin{eqnarray}
\label{eqn:mhfluor}
s_i(t) &\sim& P \bigl(F_i(t\dvtx T)|s_i(t),C_i(t);\mathbf{n}_{\setminus i}  \bigr)
\times P \bigl(s_i(t)|s_i(t-\Delta);\mathbf{n}_{\setminus i}  \bigr),
\nonumber\\[-8pt]\\[-8pt]
C_i(t) &=& C_i(t-\Delta) - \Delta/\tau^c_i \bigl(C_i(t-\Delta)-C_i^b\bigr) + A_i
n_i(t),\nonumber
\end{eqnarray}
and by appending the samples $s_i(t)$, $0<t \leq T$, we obtain a
sample from the desired density $P(\mathbf{n}_i|\mathbf{n}_{\setminus
i},\mathbf{F}_i)$.  Here the spiking term $P(s_i(t)|s_i(t-\Delta);\mathbf{n}_{\setminus i} )$ may include weak-coupling terms or truncations to
keep the state-space tractably bounded, as discussed in the previous
section.  [Again, recall the correspondence between the $K$-string
state variable $s_i(t)$ and the binary spiking variable $n_i(t)$.]

To calculate $P(F_i(t\dvtx T)|s_i(t),C_i(t))$, we can use the standard
backward HMM recursion \citet{Rabiner89},
\begin{equation}\label{eqn:genrec}
\qquad P\bigl(F_i(t\dvtx T)|s_i(t),C_i(t)\bigr)=P(F_i(t)|C_i(t))P\bigl(F_i(t+\Delta\dvtx T)|s_i(t),C_i(t)\bigr),
\end{equation}
with
\begin{eqnarray}\label{eqn:genrec1}
&&P\bigl(F_i(t+\Delta\dvtx T)|s_i(t),C_i(t)\bigr)\nonumber \\
&&\qquad =\sum _{s_i(t+\Delta)}\int dC_i(t+\Delta) P\bigl(F_i(t+\Delta\dvtx T)|s_i(t+\Delta),C_i(t+\Delta)\bigr)
\\
&&\qquad \quad\hphantom{\sum _{s_i(t+\Delta)}\int}
{}\times P\bigl(s_i(t+\Delta)|s_i(t);\mathbf{n}_{\setminus i}\bigr)P\bigl(C_i(t+\Delta)|s_i(t+\Delta),C_i(t)\bigr).\nonumber
\end{eqnarray}
We have already discussed the transition probability
$P(s_i(t+\Delta)|s_i(t);\mathbf{n}_{\setminus i})$.  The observation
density $P(F_i(t\dvtx T)|s_i(t),C_i(t))$ is a continuous function\break of~$C_i(t)$.
We could solve this backward recursion directly by
breaking the~$C_i(t)$ axis into a large number of discrete intervals
and employing standard numerical integration methods to compute the
required integrals at each time step.  However, this approach is
computationally expensive in the high SNR regime [i.e., where the
ratio of the spike-driven calcium bump $A_i$ is large relative to the
fluorescence noise scale $\sqrt{V(C_i(t))}$], where a fine
discretization becomes necessary.  \citet{Vogelstein2009} introduced a
more efficient approximate recursion for this density that we adapt
here.  The first step is to approximate $P(F_i(t\dvtx T)|s_i(t),C_i(t))$
with a mixture of Gaussians; this approximation is exact in the limit
of linear and Gaussian fluourescence observations [i.e., in the case
that the $S(\cdot)$ is linear and $V(\cdot)$ is constant in
equation~(\ref{eqn:genmodelF})], and works reasonably in practice [see
\citet {Vogelstein2009} for further details].

It is straightforward to see in this setting that we require
$2^K2^{T-t}$ mixture components to represent
$P(F_i(t\dvtx T)|s_i(t),C_i(t))$ exactly; each term in the mixture
corresponds to a distinct sequences of spikes $\mathbf{n}_i(t\dvtx T)$ and
initial conditions $s_i(t)$.  For large $T$, such a mixture of course
cannot be computed explicitly.  However, as \citet{Vogelstein2009}
pointed out, we may further approximate the intractable $2^K2^{T-t}$
mixture with a smaller $2^K(T-t+1)$ mixture, parametrized by the total
number of spikes $\sum_{t'=t}^{T}n_i(t')$ instead of the full spike
sequence $\mathbf{n}_i(t\dvtx T)$ (since mixture components with the same
total number of spikes overlap substantially, due to the long
timescale $\tau_i^c$ of the calcium decay); thus, we may avoid any
catastrophic exponential growth in the complexity of the
representation.

In order to calculate and update this approximate mixture, we proceed
as follows.  At each timestep $t$ we represent
$P(F_i(t\dvtx T)|s_i(t),C_i(t))$ as a mixture of $2^K (T-t+1)$ Gaussian
components with weights $p_{s,k}$, means $m_{s,k}$, and variances
$v_{s,k}$, for $2^K$ distinct states $s$ and $k=1,\ldots,T-t+1$.  In
order to update this mixture backward one timestep, from time $t$ to
time \mbox{$t-\Delta$}, we first integrate over $C_i(t+\Delta)$ in
equation (\ref{eqn:genrec1}).  Since $C_i(t)$ evolves deterministically
given $n_i(t)$, this reduces to updating the means and the variances
in each Gaussian component,
\begin{eqnarray}\label{step1}
m_{s,k} &\rightarrow&  [m_{s,k} - \Delta/\tau^c_i C^i_b -A_i
n_i(t)  ] /   [1-\Delta / \tau^c_i] ,\nonumber \\
v_{s,k}&\rightarrow&
v_{s,k}/(1-\Delta/\tau^c_i)^2 ,\\
p_{s,k}&\rightarrow&
p_{s,k}/(1-\Delta/\tau^c_i). \nonumber
\end{eqnarray}
Second, we perform the multiplication with the observation density
$P(F_i(t)|\allowbreak C_i(t))$ in equation (\ref{eqn:genrec}).  In the case that
$F_i(t)$ is linear and Gaussian in $C_i(t)$, each such product is
again Gaussian, and the means and the variances are updated as follows
[$M_c(t)$ and $V_c(t)$ are the mean and the variance for
$P(F_i(t)|C_i(t))$, resp.]:
\begin{eqnarray}\label{step2}
m_{s,k} &\rightarrow& \bigl(M_c(t-\Delta)v_{s,k} + m_{s,k}V_c(t-\Delta)\bigr)/\bigl(v_{s,k} + V_c(t-\Delta)\bigr) \nonumber \\
v_{s,k} &\rightarrow& V_c(t-\Delta)v_{s,k}/\bigl(v_{s,k}+V_c(t-\Delta)\bigr) \nonumber\\
p_{s,k} &\rightarrow& p_{s,k}\times \bigl(2 \pi
\bigl(v_{s,k}+V_c(t-\Delta)\bigr)\bigr)^{-1/2}\\
&&{}\times\exp \biggl[-\frac{M_c(t-\Delta)^2}{2V_c(t-\Delta)}
-\frac{m_{s,k}^2}{2v_{s,k}}\nonumber \\
&&\hphantom{{}\times\exp \biggl[}
{}+\frac{(M_c(t-\Delta)v_{s,k} + m_{s,k}V_c(t-\Delta))^2}
{2(v_{s,k} + V_c(t-\Delta)) V_c(t-\Delta)v_{s,k}}
 \biggr].\nonumber
\end{eqnarray}
More generally [i.e., in the case of a nonlinear $S(\cdot)$ or nonconstant
$V(\cdot)$ in equation~(\ref{eqn:genmodelF})], standard Gaussian approximations
may be used to arrive at a similar update rule; again, see
\citet{Vogelstein2009} for details.

Third, we perform the summation over $s_i(t+\Delta)$ in
equation~(\ref{eqn:genrec1}), and reorganize the obtained mixture to reduce
the number of components from $2(T-t+1)$ to $(T-t+2)$ for each
$s_i(t)$.  Equation (\ref{eqn:genrec1}) doubles each mixture component into
two new Gaussians, corresponding to the two terms in the sum for
$n_i(t+\Delta)=1$ or $n_i(t+\Delta)=0$.  For brevity, we denote these
terms as $w^+=P(s^+(t+\Delta)|s_i(t))$ and
$w^-=P(s^-(t+\Delta)|s_i(t))$, where $s^+(t+\Delta)$ stands for the
state $s_i(t+\Delta)$ describing a spike at time $t+\Delta$, and
$s^-(t+\Delta)$ denotes the absence of a spike at time $t+\Delta$.  We
group all new components in pairs such that each pair corresponds to a
given number of spikes on the interval $t\dvtx T$.  Each such pair of
Gaussian components is then merged into a~single Gaussian with
equivalent weight, mean, and variance:
\begin{eqnarray}\label{step3}
p_{s,k}&\rightarrow& w^+ + w^-, \nonumber \\
m_{s,k}&\rightarrow&(w^+ m_{s^+,k-1} + w^- m_{s^-,k})/(w^+ + w^-), \\
v_{s,k}&\rightarrow& v_{s,k}^m + (w^+ v_{s^+,k-1}+ w^- v_{s^-,k})/(w^+ + w^-),\nonumber
\end{eqnarray}
and the $v_{s,k}^m$ term corresponds to the variance of the means,
\begin{eqnarray}
v_{s,k}^m &=& \bigl(w^+ \bigl(m_{s^+,k-1}(t) - m_{s,k}(t-\Delta)\bigr)^2\nonumber\\[-8pt]\\[-8pt]
&&{}+ w^- \bigl(m_{s^-,k}(t) - m_{s,k}(t-\Delta)\bigr)^2\bigr)/(w^+ + w^-).\nonumber
\end{eqnarray}
See Figure~\ref{fig:fluorrecursion} below for an illustration.

\subsection{Simulating populations of spiking neurons}\label{secIIF}
To test the performance of different sampling algorithms, we simulated
a population of $N=50$--800 neurons, following the approach described
in \citet{Mishchenko2010}.  Briefly, we simulated a spontaneously
active randomly connected neural network, with each neuron described
by model equation (\ref{eqn:genmodel}), and connectivity and functional
parameters of individual neurons chosen randomly from distributions
based on the experimental data available for cortical networks in the
literature [\citet{Sayer1990}; \citet{Braitenberg1998}; \citet{Urquijo2000}; \citet{Lefort2009}].
Networks consisted of 80\% excitatory and 20\% inhibitory neurons
[\citet{Braitenberg1998}; \citet{Urquijo2000}]. Neurons were connected to each
other in a sparse, spatially homogeneous manner: the probability that
any two neurons $i$ and $j$ were connected (i.e., that either $w_{ij}$
or $w_{ji}$ was nonzero) was $0.1$ [\citet{Braitenberg1998}; \citet{Lefort2009}].
The scale of the connectivity weights $w_{ij}$ was matched to results
from the cortical literature, as cited above, and the overall average
firing rate of the networks was set to be about 5 Hz.  The
connectivity waveforms $w_{ij}(t)$ were modeled as exponential
functions with time constant fixed for all neurons at $10$ msec; for
the self-coupling terms $w_{ii}(t)$, neurons strongly inhibited
themselves over short time scales (an absolute refractory effect of 2
ms) and weakly inhibited themselves with an exponentially-decaying
weight over a timescale of 10 ms.  Finally, we used an exponential
nonlinearity, $f(\cdot)=\exp(\cdot)$, for simplicity.  Again, see
\citet{Mishchenko2010} for full details and further
discussion.\looseness=-1

\section{Results}\label{res}

The efficiency of any Metropolis--Hastings sampler is determined by the
ease with which we can sample from the proposal density (and compute
the acceptance probability) versus the degree to which the proposal
approximates the true target $P(\mathbf{n}_i|\mathbf{n}_{\mathit{observed}})$.  In
this section we will numerically compare the efficiency of the
proposal densities we have discussed above.  In particular, we will
examine the following proposal densities, listed in rough order of
complexity:
\begin{itemize}
\item Time-homogeneous Poisson process.
\item Point process with log-conditional intensity function given by
  the delayed input $J^-_i(t)$, equation (\ref{eqn:lagcur}).
\item Point process with log-rate determined by full effective input in
  weak-coupling approximation $\tilde J_i(t)$,
  equation (\ref{eqn:weakcur}).
\item Hybrid truncated HMM proposal including weak-coupling terms [equation~(\ref{HMMMH6})].
\end{itemize}
As a benchmark, we also compare these samplers against a simple
pointwise Gibbs sampler, in which we draw from
$P(n_i(t)|\mathbf{n}_{i,\setminus t},\mathbf{n}_{\setminus i})$ sequentially over~$t$.

We begin by inspecting a simpler toy model of neural spiking.  In this
model both refractory and interneuronal coupling effects are assumed
to be short, that is, on the time scale of $\approx $20 msec.  This
simplification allows us to characterize the neural state fully by
$K\approx 10$ past time bins ($\Delta =2$ ms in these simulations),
with the state variable $s_i(t)=\{n_i(t'),t-K\Delta < t' \leq t \}$.
In this case, equation~(\ref{eqn:genmodel}) describes a hidden Markov model
with a~state space that is small enough to sample from directly, using
the forward--backward procedure detailed in Section \ref{secIIB}.
Thus, in this case we can obtain the probability distribution
$P(\mathbf{n}_i|\mathbf{n}_{\mathit{observed}})$ explicitly, and compare different MH
algorithms against a ground truth.

\begin{figure}

\includegraphics{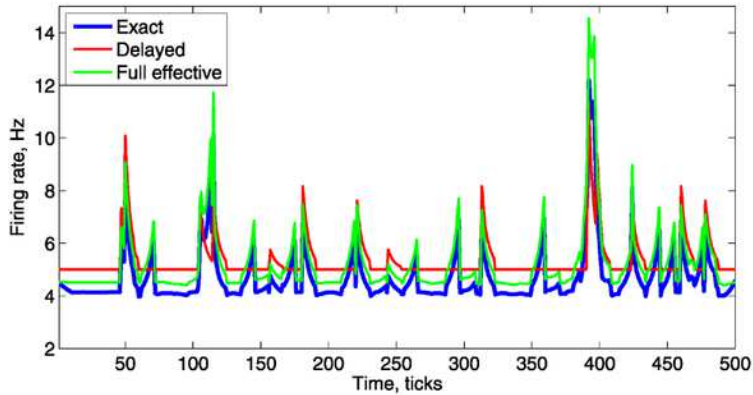}

\caption{Quantifying the approximation accuracy of two spike train
proposal densities.  Blue trace: the true spiking rate $r_i(t)$ of one
neuron conditioned on the spiking activity of all other neurons in a
population of $N=50$ neurons; the rate is computed exactly via
forward--backward procedure described in Section \protect\ref{secIIB}.  A
population of spontaneously spiking neurons with neurophysiologically
feasible parameters (Section \protect\ref{secIIF}) was simulated for a total
of $1$ sec at a time resolution of $\Delta=2$~msec.  Approximate
spiking rates obtained using just the delayed input, $J^-(t)$
[equation~(\protect\ref{eqn:lagcur})], or full weak-coupling input, $\tilde J(t)$
[equation~(\protect\ref{eqn:weakcursimple})], are shown in red and green,
respectively.  (In particular, the delayed input~$J^-_i(t)$ and the full
effective input $\tilde J_i(t)$ are computed as described in Section
\protect\ref{secIIC}, and then we approximate the rates using $r_i(t) =
f[J_i^-(t)]$ or $f[\tilde J_i(t)]$.)  While the proposal based on the
delayed inputs $J^-(t)$ reproduces the true spiking rate somewhat
poorly, the weak-coupling~$\tilde J(t)$ approximation is significantly
more accurate.}
\label{fig:rates}
\end{figure}

In Figure \ref{fig:rates} we inspect the true instantaneous posterior
spiking rate $r_i(t)=P(n_i(t)|\mathbf{n}_{\mathit{observed}})/\Delta$ for the
hidden neuron in this toy model, calculated exactly with the
forward--backward procedure and estimated using different effective
rates for a few of the MH proposals discussed above.  We see that the
true rate $r_i(t)$ varies widely around its mean value, implying that
the simplest proposal density (the time-homogeneous Poisson process)
will result in rather low acceptance rates, as indeed we will see
below.  Similarly, a naive approximation for $r_i(t)$ using only the
delayed input $J_i^-(t)$ fails to capture much of the structure of
$r_i(t)$.  Incorporating both the past and the future spiking activity
of the observed neurons via the weak-coupling input $\tilde J_i(t)$
leads to a much more accurate approximation of the true rate $r_i(t)$.

\begin{figure}

\includegraphics{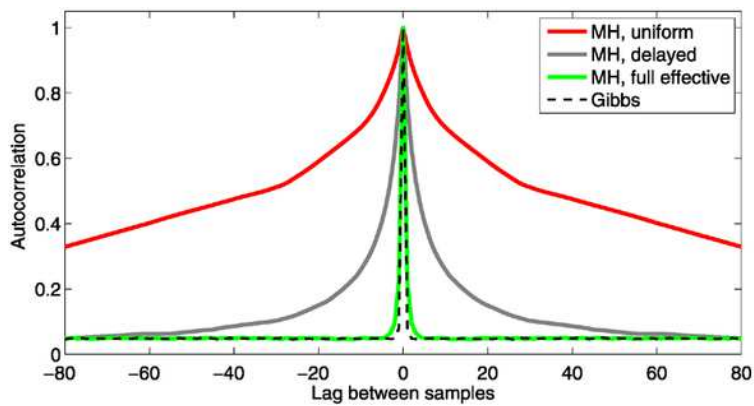}

\caption{Autocorrelation functions for Gibbs and MH samplers.  (The
autocorrelation function is averaged over all time bins in the spike
train.)  Echoing the results of Figure \protect\ref{fig:rates}, we see that the
homogeneous (``uniform'') Poisson MH algorithm mixes slowly; the
sampler based on the full weak-coupling input $\tilde J_i(t)$ mixes
significantly more quickly than the sampler based on the delayed
inputs $J^-_i(t)$.  The standard Gibbs sampler performed about as well
as the weak-coupling MH sampler.  Simulation details are as in
Figure \protect\ref{fig:rates}, except a~$10$~sec spike train was simulated
here to collect a~sufficient amount of data to distinguish the
different algorithms.}
\label{fig:aucorrelations}
\end{figure}

\begin{figure}

\includegraphics{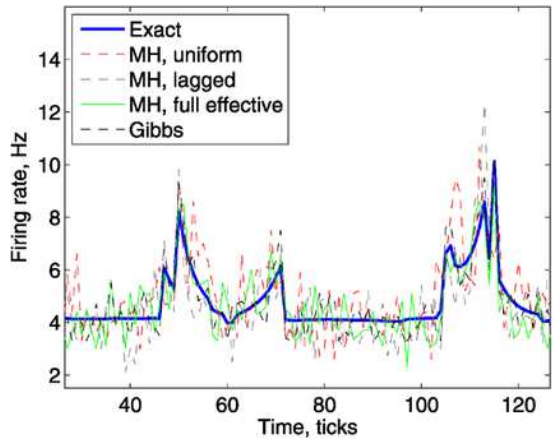}

\caption{Comparing the accuracy of MH and Gibbs samplers.  Each trace
indicates the posterior spiking rate $r_i(t)$ for one hidden neuron,
estimated from 5,000 samples using the Gibbs sampler and the three MH
samplers compared in the preceding two figures.  The results are
similar: the homogeneous Poisson proposal performs badly [in the sense
that the $r_i(t)$ computed based on these $5\mbox{,}000$ samples approximates
the true $r_i(t)$ poorly], while the weak-coupling and Gibbs samplers
outperform the sampler based on the delayed inputs~$J^-_i(t)$ (in
terms of variance around the true rate, computed via the full
forward--backward HMM method, shown in blue).  MH acceptance rates,
$R$, varied from $R\approx 0.75$ for~MH using the homogeneous Poisson
proposal to $R\approx 0.98$ for MH using the proposal with full
effective input~$\tilde J_i(t)$.  Simulation details are as in
Figure \protect\ref{fig:rates}.  A shorter time interval of the simulated spike
train is shown for greater clarity here.}
\label{fig:marginals}
\end{figure}

Similar results are obtained when we apply the MH algorithm using
these proposal densities (Figures \ref{fig:aucorrelations} and
\ref{fig:marginals}).  We use MH with $M=5\mbox{,}000$ samples for each
proposal density, with a burn-in period of 1,000 samples, to obtain
both the autocorrelation functions (Figure~\ref{fig:aucorrelations}) and
estimates for the instantaneous spiking rate $r_i(t)$
(Figure~\ref{fig:marginals}).  We observe that the homogeneous Poisson
proposal density leads to a very long autocorrelation scale, implying
inefficient mixing, that is, long runs are necessary to obtain accurate
estimates for quantities of interest such as $r_i(t)$.  Indeed, we see
in Figure~\ref{fig:marginals} that 5,000 samples are insufficient to
accurately reconstruct the desired rate $r_i(t)$.  Similarly, for the
proposal density based on the delayed inputs $J^-(t)$, the
autocorrelation scale is shorter, but still in the range of 10--20
samples.  The weak-coupling proposal, which incorporates information
from both past and future observed spiking activity, mixes quite well,
with an autocorrelation scale on the order of a single sample, and
leads to an accurate reconstruction of the true rate in
Figure~\ref{fig:marginals}.  Interestingly, the simplest Gibbs sampling
algorithm also performs well, with an autocorrelation length similar
to that of the best MH sampler shown here.

\begin{figure}

\includegraphics{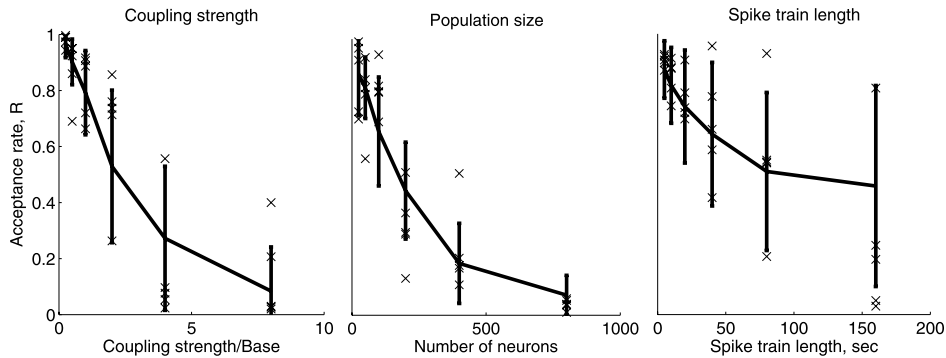}

\caption{Acceptance rate of the weak-coupling MH algorithm as a
function of coupling strength, $C$, neural population size, $N$, and
spike length, $T$. We simulated sparsely connected networks of
inhibitory and excitatory neurons as described in Section
\protect\ref{secIIF}, with refractory effects up to $10$ msec long and
interneural interactions up to $50$ msec long.  Here a coupling
strength value of $C=1$ corresponds to the neurobiologically motivated
set of parameters in Section \protect\ref{secIIF} (also in
Figures~\protect\ref{fig:rates}--\protect\ref{fig:marginals}); other values of $C$
corresponded to scaling $f(J(t))\rightarrow f(C\cdot J(t))$ in
equation~(\protect\ref{eqn:genmodel}).  $10$ sec of neural activity was simulated
at a time resolution of $\Delta = 2$ msec, with the neural population
spiking at $\approx $4--5~Hz.  64 trials of 500 samples were simulated,
with a new random neural network generated in each trial, from which
we estimated $P(\mathbf{n}_i|\mathbf{n}_{\setminus i};\mathbf{w})$.  Average
acceptance rate $R$ and standard deviation are shown for such trials, as
well as examples of several individual trials (``x'').  MH algorithm
performance degraded significantly as $C$ or $N$ increased.  On the
other hand, for larger values of $T$ performance degraded much less
substantially, and even for the largest $T$ we examined ($160$ sec)
the acceptance rate remained above $R\sim 0.4$.}
\label{fig:combinedMH}
\vspace*{6pt}
\end{figure}

We also study the performance of the weak-coupling proposal as a
function of the strength of the interneuronal interactions in the
network, and as a~function of the number of neurons $N$ in the
population and the length~$T$ of the desired spike train
(Figure~\ref{fig:combinedMH}).  The acceptance rate falls at a rate
approximately inverse to the coupling strength, which seems sensible,
since this proposal is based on the approximation that the coupling
strength is weak.  We observe similar behavior with respect to the
size of the neural population. In particular, the acceptance rate
drops below $R \sim 0.1$ when $N\sim 600$--1,000.  On the other hand,
increasing the spike train length $T$ affects performance to only a
moderate degree: even when the length of the sampled spike train is
increased from $10$ sec to $160$ sec, the acceptance rate remains
above $R \sim 0.4$.

\begin{figure}

\includegraphics{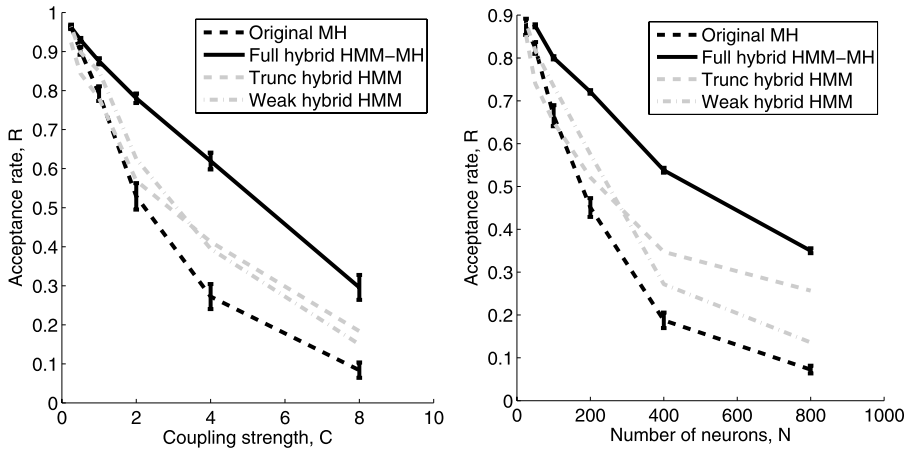}

\caption{Acceptance rate of hybrid HMM-MH algorithm
[equation~(\protect\ref{HMMMH6})]
as a function of the overall coupling strength $C$ and neural
population size $N$.  The hybrid algorithm (solid black line) performs
substantially better than the original MH algorithm described in
Section \protect\ref{secIIC} (dashed black line), and uniformly better than
alternative hybrid strategies, for example, where only a short-scale
truncated HMM is used or where all interneural interactions are
accounted for via weak-coupling approximation (gray lines).
Simulations are as in Figure~\protect\ref{fig:combinedMH}.}
\label{fig:combinedhyHMMMH}
\end{figure}

One would expect that the performance of the MH algorithm in the case
of more strongly-coupled neural networks can be improved by including
strong short-term interaction and refractory effects explicitly into
the proposal density.  We discussed such an approach in Section
\ref{secIID}.  For weakly coupled networks, we expect the performance
of this hybrid algorithm to be similar to that of our original
proposal density; however, for more strongly coupled neural networks,
we expect a better performance.  This expectation is borne out in the
simulations shown in Figure \ref{fig:combinedhyHMMMH}: the hybrid
sampler (solid black line) performs significantly better than the
original MH algorithm (dashed black line) and uniformly better than
alternative hybrid strategies, for example, where only the short-scale
truncated HMM is used, or where all interneural interactions are
accounted for via the weak-coupling approximation, while the
self-terms $w_{ii}$ are accounted for via a truncated HMM (gray
lines).  Similar results are obtained when we vary the size of the
neural population~$N$ (Figure \ref{fig:combinedhyHMMMH}).

In the simulation settings described above (timestep $\Delta=2$ msec,
with coupling currents $w_{ij}$ lasting up to $50$ ms) the MH
algorithm (coded in Matlab without any particular algorithmic
optimization) took $\approx$300--400 sec to produce 1,000 samples of
1,000 time-ticks each on a PC laptop (Intel Core Duo 2 GHz), dominated
by the time necessary to produce 1,000 spike train proposals using the
forward recursion.  The hybrid algorithm had a~similar computational
complexity ($t_{\max}=10$ ms in the truncated HMM).  The Gibbs
algorithm had a somewhat higher computational cost, due largely to the
fact that updates of the currents $J_i(t)$ for up to $50$ ms needed to
be performed during each step, to decide whether to flip the state of
each spike variable~$n_i(t)$.  This resulted in running times for the
Gibbs sampler which were 5--20 times slower than for the MH sampler.
The Gibbs updates are even slower in the calcium imaging setting
(discussed at more length in the next section), due to the fact that
each update requires us to update $C_i(t)$ over $n_c \tau^c_i$
timesteps (recall Section \ref{sec:sampl-ca}) for each proposed flip
of $n_i(t)$.

\subsection{Incorporating calcium fluorescence imaging observations}

\begin{figure}

\includegraphics{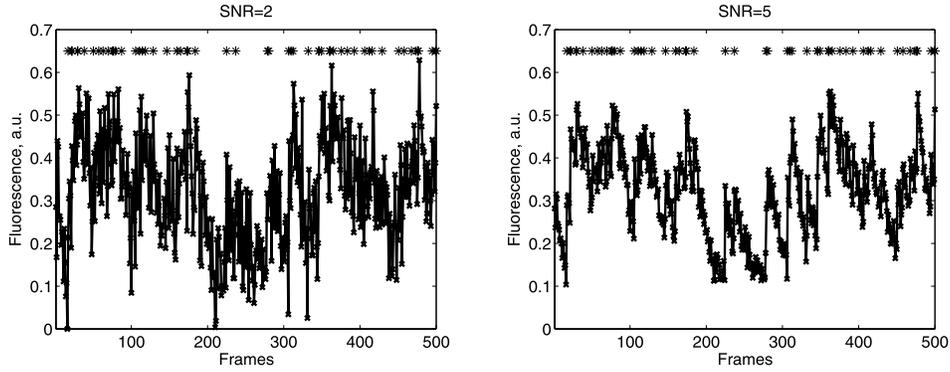}

\caption{Example of observed fluorescence for low $\mathit{SNR}\approx 2$ and
relatively high $\mathit{SNR}\approx 5$.  $\Delta=2$~msec, and calcium imaging
frame-rate $\mathit{FR}=50$ Hz. $10$ sec spike trains were simulated, with the
neural population spiking at $\approx $4--5 Hz. Actual spikes of the
target neuron are indicated with stars.}
\label{fig:fluortrace}
\vspace*{6pt}
\end{figure}

\begin{figure}[b]

\includegraphics{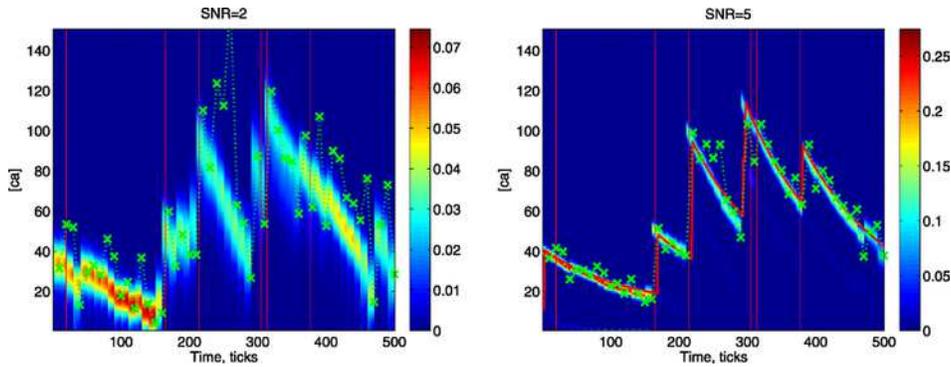}

\caption{Example of $P(\mathbf{F}_i(t\dvtx T)|C_i(t))$ calculated as a
$T$-mixture of Gaussians, as a function of time (x-axis) and $C_i(t)$
(y-axis); colorbar indicates the probability density at time $t$.  For
reference, true calcium concentration is shown in red, and ``observed''
calcium concentration at the imaging frames [i.e., $S^{-1}(F_i(t))$] is
shown with green ``x.''  Simulation details are as in
Figure \protect\ref{fig:fluortrace}, first second (50 frames) is shown for clarity.
Time ticks correspond to $\Delta=2$ msec.
Actual spike times of the target neuron
are shown with red lines.}
\label{fig:fluorrecursion}
\end{figure}

\begin{figure}

\includegraphics{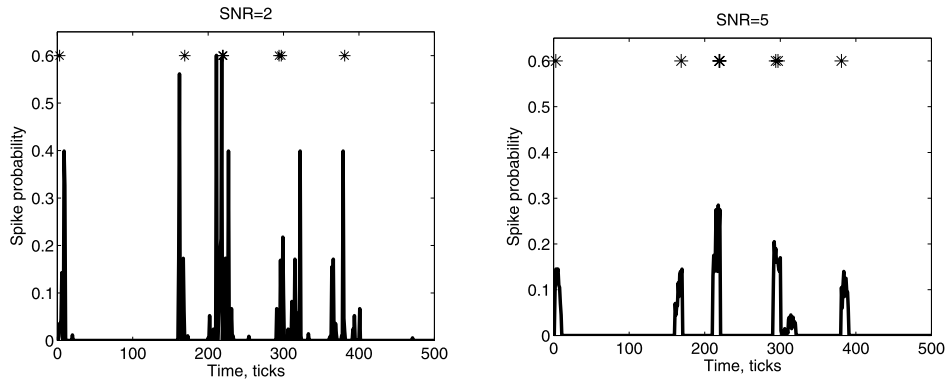}

\caption{Posterior spike probability, $P(n_i(t)|\mathbf{n}_{\setminus
i},\mathbf{F}_i)$, estimated using MH algorithm. While the MH algorithm
performs well for $\mathit{SNR}=5$ and above ($R\approx 0.8$), for low $\mathit{SNR}=2$
acceptance rate is only $R\approx 0.01$.  Simulation details are as in
Figure \protect\ref{fig:fluortrace}, first second (50 frames) is shown for clarity.
Time ticks correspond to $\Delta=2$ msec.
Actual spikes of the target neuron are
shown with asterisks.  Note that in the high-SNR case the sampler
successfully recovers the smoothly-varying $P(n_i(t)|\mathbf{n}_{\setminus i},\mathbf{F}_i)$
on a finer time-scale ($\Delta\approx 2$
msec) than the original fluorescence imaging data provided
($\Delta_{\mathit{FR}}\approx 20$~msec).  However, in the low-SNR case the
recovered firing rate is overly spiky and variable due to the slow
mixing speed of the M-H chain; recall that similar behavior is visible
in Figure~\protect\ref{fig:marginals}.}
\label{fig:fluorsamples}
\end{figure}

Next we examine the performance of the method we developed in Section
\ref{sec:sampl-ca} for sampling from $P(\mathbf{n}_i|\mathbf{n}_{\setminus
i},\mathbf{F}_i)$.  We find that this sampler performs quite well in
moderate and high SNR settings.  More precisely, for values of eSNR
greater than $\approx $5 [recall the definition of eSNR in
equation~(\ref{eq:eSNR})], the conditional distribution $P(\mathbf{n}_i | \mathbf{F}_i)$
is localized near the true spike train quite effectively,
leading to a high MH acceptance rate (Figures
\ref{fig:fluortrace}--\ref{fig:fluorsamples}).  Indeed, we find [as in
\citet{Vogelstein2009}] that it is possible to achieve a sort of
``super-resolution'' in the sense that the MH algorithm can
successfully return $P(\mathbf{n}_i | \mathbf{F}_i)$ on the time-scale of
$\Delta=2$ msec even when fluorescence observations are obtained at a
much lower frame-rate (here $\mathit{FR}=50$ Hz).

\begin{figure}

\includegraphics{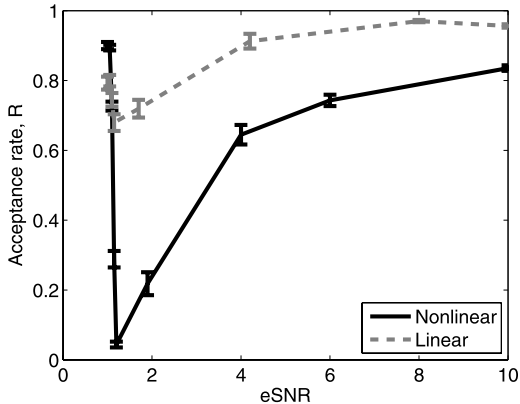}

\caption{Acceptance rate of MH algorithms as a function of the calcium
imaging effective SNR, equation~(\protect\ref{eq:eSNR}).  Calculations for two
calcium signal models, with a toy linear (gray dashed) and realistic
Hill (solid black) transfer function, $S(C_i(t))$, are shown.  [See
Vogelstein et~al. (\protect\citeyear{Vogelstein2009}); Mishchenko, Vogelstein and
  Paninski (\protect\citeyear{Mishchenko2010}) for further details on the
precise form of the nonlinear function $S(C_i(t))$; in this case,
three or four spikes were sufficient to drive the calcium variable
into a regime where the fluorescence signal was significantly
saturated.]  Performance of the MH algorithm is good both for high and low
eSNR, but can suffer for intermediate $\mathit{eSNR}\approx 1.1$--3.  This
performance drop is primarily due to deviations of the conditional
distribution $P(C_i(t)|F_i(t))$ from the Gaussian shape, as
exemplified by the much better performance in the model where $S(\cdot)$
is linear and $P(C_i(t)|F_i(t))$ is thus Gaussian.  Simulation details
are as in Figure~\protect\ref{fig:fluortrace}.  For each eSNR 50 simulations for
different neural populations were performed, and the average and
standard error of these simulations are shown. }
\label{fig:perffluor}
\end{figure}

Conversely, for low values of $e\mathit{SNR}$, for example, $e\mathit{SNR} \approx 1$, the
backward density $P(\mathbf{F}_i(t\dvtx T)|n_i(t),C_i(t))$ becomes
noninformative [i.e., relatively flat as a function of $n_i(t)$ and
$C_i(t)$], and the acceptance rate of our MH algorithm reverts to the
rate obtained under the conditions of no calcium imaging data.
However, in the low-to-moderate SNR regime (e.g., $e\mathit{SNR} \approx 2$),
the performance of the MH sampler can drop substantially.  This is
primarily due to deviations in the shape of $P(F_i(t)|C_i(t))$ from
Gaussian at low SNR.  Recall that we made several approximations in
computing the backward density: first, this density is truly a
mixture of ${\sim} 2^{T-t}$ components, whereas we approximate it with a
mixture of only ${\sim} T-t$ components.  Second, we assume that each
mixture component is Gaussian.  Although the fluorescence~$F_i(t)$ is
described by normal statistics given the calcium variable $C_i(t)$,
the relationship between the fluorescence mean and calcium transient
can be nonlinear, and the variance may depend on $C_i(t)$ [recall
equation~(\ref{eqn:genmodelF})].  This makes the conditional distribution of
$C_i(t)$ non-Gaussian in general, particularly at low-to-moderate
levels of SNR (where the likelihood term is informative but not
sufficiently sharp to justify a simple Laplace approximation).  We
tested the impact of each of these approximations individually by
constructing a set of toy models where these different approximations
were made exact; we found that the non-Gaussianity of
$P(C_i(t)|F_i(t))$, due to nonlinear dependence of $F_i(t)$ on
$C_i(t)$, was the primary factor responsible for the drop in
performance.  Thus, we expect that in cases where the saturating
function $S(\cdot)$ is close to linear, the sampler should perform well
across the full SNR range (Figure~\ref{fig:perffluor}); in highly
nonlinear settings, more sophisticated approximations [based on
numerical integration techniques such as expectation propagation
\citet{MINKAPHD}] may be necessary, as discussed in the \hyperref[sec:discussion]{Conclusion}
section below.  The MH sampler here took about twice as long as in the
fully-observed spike train case discussed in the previous section,
largely due to the increased complexity of the backward recursion
described in Section \ref{sec:sampl-ca}.

\begin{figure}

\includegraphics{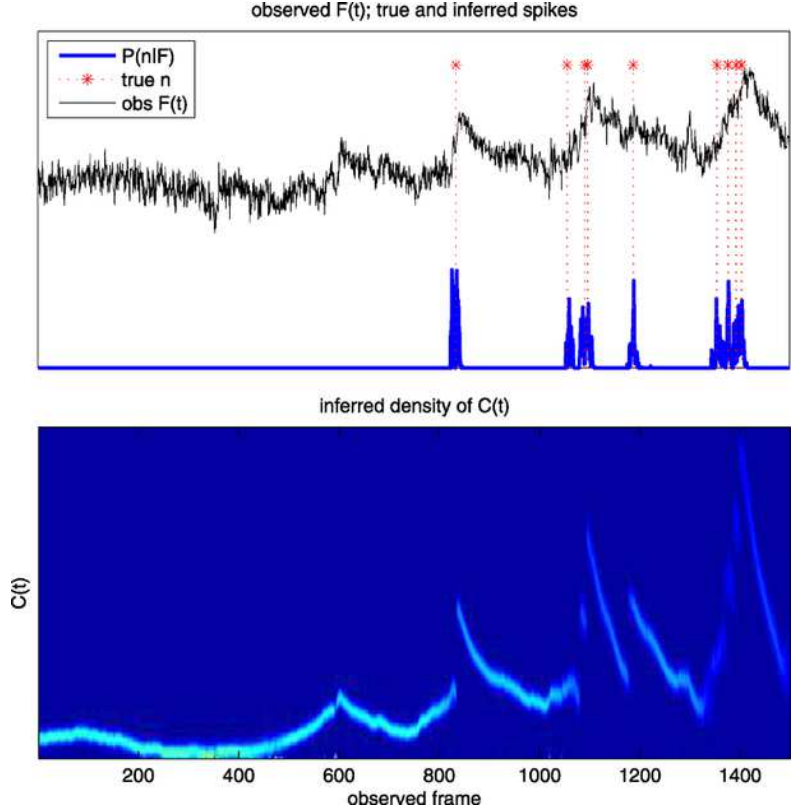}

\caption{Application to real data (sample data courtesy of T.\ Sippy,
  R.\ Yuste, J.\ Vogelstein).  Top: observed $F(t)$ (black) and true
  spike times (red) were recorded from a~single neuron via
  simultaneous fluorescence imaging and intracellular patch-clamp
  electrophysiological recording; see Vogelstein et~al. (\protect\citeyear{Vogelstein2009}) for
  further experimental details.  Blue trace indicates posterior
  $P(n(t)|\mathbf{F})$ computed by the hybrid sampler.  Bottom:
  $p(C(t)|\mathbf{F})$ computed by the hybrid sampler.  Y-axis units
  are arbitrary in this case and have been suppressed.  Note that the
  sampler infers spikes and jumps in $C(t)$ at the correct times.}
\label{fig:real}
\end{figure}

Given the good performance of Gibbs sampling noted in the calcium-free
setting discussed above, we also examined the Gibbs approach here.
However, we found that the Gibbs algorithm was not able to procure the
samples successfully given calcium imaging observations.  In many
cases we found that the Gibbs sampler converged rapidly to a
particular spike train close to the truth, but would then become
``stuck.''  If initialized again, the sampler would often converge to
a different spike train, close to the truth, only to become stuck
again.  This behavior is due to the fact that the conditional
distributions $P(n_i(t)|\mathbf{n}_{i,\setminus t},\mathbf{F}_i)$ can be
quite sharply concentrated, leading to poor mixing between spike train
configurations with high posterior probability; of course, this is a
common problem with the Gibbs sampler (and indeed, this
well-understood poor mixing behavior of the standard Gibbs chain is
what led us to develop the more involved methods presented here in the
first place).  Roughly speaking, the extra constraints imposed by the
fluorescence observations in $P(n_i(t)|\mathbf{n}_{i,\setminus t},\mathbf{F}_i)$ relative to $P(n_i(t)|\mathbf{n}_{i,\setminus t})$ make the former
distribution more ``frustrated,'' in physics language, making it
harder for the Gibbs sampler to reach nearby states with high
posterior probability and leading to the relatively slower Gibbs
mixing rate in the calcium-imaging setting.

Finally, we applied the hybrid sampler to a sample of real calcium
fluorescence imaging data (Figure~\ref{fig:real}), in which $F(t)$ and
the true spike times~$n(t)$ were recorded simultaneously.  Thus, we
have access to ground truth for $n(t)$ here, though of course only the
observed fluorescence $\mathbf{F}$ is used to infer
$p(n(t)|\mathbf{F})$.  The model parameters were estimated using the
EM method discussed in \citet{Vogelstein2009}; again, only the observed
$\mathbf{F}$ was used to infer the model parameters, not the true
spike times $n(t)$.  About $50$ spikes' worth ($5\mbox{,}000$ fluorescence
frames) of data was sufficient to adequately constrain the parameters.
Then we applied the hybrid sampler using these parameters to the
subset of data shown in Figure~\ref{fig:real}.  The sampler does a
good job of recovering the spike rate $n(t)$ from the observed
fluorescence data $\mathbf{F}$, and seems to do a reasonable job of
recovering the corresponding jumps in $p(C(t)|\mathbf{F})$ that occur
at spike times.  Note that we do not have access to the true
intracellular calcium concentration $C(t)$, and therefore no ground
truth comparisons are possible for this variable.

\section{Conclusion}
\label{sec:discussion}

In this work we developed several Metropolis--Hastings approaches for
sampling from the conditional distribution of neuronal spike trains,
given either the activity of other neurons in the network or
calcium-sensitive imaging observations.  The most effective approach
was the hybrid method described in Section \ref{secIID}, which takes
advantage of the fact that strong short-term temporal dependencies
within a single spike train may be handled via forward--backward hidden
Markov model methods, while weaker long-term dependencies between
neurons may be handled with the weak-coupling expansion developed in
Section~\ref{secIIC}.  In each case, to sample efficiently from the
spike train at time $t$, it is important to incorporate not only past
but also future information (i.e., spiking observations from times both
before and after $t$); \citet{PL07} made a similar point.  In the
appendix we show that these methods may be extended rather easily to
other exponential families (not just the Bernoulli and Poisson cases
of most interest in the neuroscience setting); further applications to
weakly-coupled Markov chains in nonneural settings seem worth
exploring.

Two major avenues are open for future work.  First, as noted in
Figure~\ref{fig:fluorrecursion}, the proposed sampler suffers somewhat
in the case of strongly nonlinear fluorescence observations, largely
because in this case our mixture-of-Gaussians approximation of the
backward density $P[\mathbf{F}_i(t\dvtx T) | C_i(t)]$ can break down.  More
sophisticated methods for approximating this density are available,
and should be explored more thoroughly.  Second, as discussed in
\citet{VogelsteinPoster10}, applications of these methods to real data
are ongoing, via Monte Carlo-Expectation-Maximization methods similar
to those discussed in \citet{Vogelstein2009}; \citet{Mishchenko2010}, with the
fast sampler introduced here replacing the slower Monte Carlo
approaches discussed in \citet{Mishchenko2010}.  Calcium-fluorescence
imaging methods have exploded in popularity over the last several
years, and we hope the methods presented here will prove useful in
quantifying the cross-correlations and effective connectivity in
neural populations observed via fluorescence imaging and
multielectrode recording methods.

\begin{appendix}
\section*{Appendix: Sampling from a weakly-coupled exponential~family}
\label{sec:append-sampl-from}

As mentioned in Section \ref{secIIC}, it is straightforward to develop
a first-order proposal density in the weak-coupling limit more
generally, in the case that the variables of interest are drawn from
an exponential family distribution with linear sufficient statistics.
We begin by writing down our exponential family model, using slightly
more compact notation than in Section~\ref{secIIC}:
\begin{equation}
  \log p(\{ n_{it}\}) = \sum_{it} f(J_{it}) k(n_{it}) + g(J_{it}) + h(n_{it}),
\end{equation}
with the coupling introduced via
\begin{equation}
  J_{it} = b_{it} + \sum_{s>0, j} w^{ij}_s n_{j,t-s}.
\end{equation}
As before, we may easily expand the joint log-density to the first order:
\begin{eqnarray}
  \log p(\{ n_{it}\}) &=& \sum_{t} f(J_{it}) k(n_{it}) + g(J_{it}) +
  h(n_{it}) \\
  &&{}+ \sum_{t,i\neq j}  [f'(b_{jt}) k(n_{jt}) +
  g'(b_{jt})  ] \sum_{s>0} w_s^{ji} n_{i,t-s}\nonumber\\[-8pt]\\[-8pt]
  &&{} + \mbox{const}(n_{it}) + o(w).\nonumber
\end{eqnarray}
Now if we introduce the assumption that the sufficient statistic is
linear, that is, $k(n)=n$, then after rearranging the double sum over $s$
and $t$ we obtain
\begin{eqnarray*}
  \log p(\{ n_{it}\}) &=& \sum_{t}  \biggl[ f(J_{it}) + \sum_{s>0,i\neq
  j} w_s^{ji}  [f'(b_{j,t+s}) n_{j,t+s} + g'(b_{j,t+s})  ]
   \biggr] n_{it}\\
   &&{} + h(n_{it}) + \mbox{const}(n_{it}) + o(w),
\end{eqnarray*}
where again we have suppressed terms [such as $g(J_{it})$] which do
not invol\-ve~$n_{it}$; thus, to first order, the conditional
distribution of $\{ n_{it}\}$ given~$\{ n_{jt}\}$, \mbox{$i \neq j$}, remains
within the same exponential family, but with a parameter shift
\begin{equation}
  f(J_{it}) \to f(J_{it}) + \sum_{s>0,i\neq j} w_s^{ji}  [
f'(b_{j,t+s}) n_{j,t+s} + g'(b_{j,t+s})  ].
\end{equation}
In the canonical parameterization, $f(J)=J$, standard exponential
family theory [\citet{CasellaBerger}] shows that $g'(b_{j,t+s}) = -
E(n_{j,t+s}|b_{j,t+s})$, and the parameter shift simplifies to
\begin{equation}
  J_{it} \to J_{it} + \sum_{s>0,i\neq j} w_s^{ij}  [n_{j,t+s} -
E(n_{j,t+s}|b_{j,t+s})  ].
\end{equation}
See \citet{Beck07} for a discussion of some related results.

As a concrete example, consider the Gaussian case:
\begin{equation}
  \log p(\{ n_{it}\}) = - \sum_{it} \frac {1} {2 \sigma^2_i} (n_{it} -
  J_{it})^2 + \mbox{const.},
\end{equation}
with $J_{it}$ as above.  In this case we may define $f_i(J_{it}) =
J_{it} / \sigma^2_i$ and $g_i(J_{it}) = -J_{it}^2 / 2\sigma^2_i$; thus,
we find that the parameter shift in this case is
\begin{equation}
  f(J_{it}) = \frac {J_{it}} {\sigma^2_i} \to \frac {J_{it}}
{\sigma^2_i} + \sum_{s>0,i\neq j} \frac {w_s^{ij}} {\sigma^2_j}  [
n_{j,t+s} - b_{j,t+s}  ].
\label{eq:gauss-shift}
\end{equation}
In this linear-Gaussian case, we may compute the exact conditional
distribution of $n_i$ via the usual Gaussian conditioning formula; the
necessary covariance and inverse covariance matrices may be obtained
via standard AR model computations.  It is straightforward to check
that this exact formula agrees with equation~(\ref{eq:gauss-shift}) up to
$o(w)$ terms in the small-$w$ limit.
\end{appendix}

\section*{Acknowledgments}
Thanks to T.\ Machado, M.\ Nikitchenko, J.\ Pillow and J.~Vogelstein
for many helpful conversations, and again to T.\ Sippy and R.\ Yuste
for the data example shown in Figure~\ref{fig:real}.    We also gratefully acknowledge the use of
the Hotfoot shared cluster computer at Columbia University.


\printaddresses

\end{document}